\documentstyle{article}
\def\graphic #1#2#3#4#5{

    \noindent
    \centerline{\hrulefill}
    \leftline{\hbox to#1{\special{anisoscale #3, #1 #2}\hfil}}
    \vspace*{#2} \relax
    \vskip -3.9 cm
    \hskip 4.8 cm
    {\large \bf Universidade do Estado do Rio de Janeiro }
    \newline

    \vskip -0.25 cm
    \hskip 7.5 cm
    {\large \bf Instituto de F{\'\i}sica }

    \vskip 1 cm
    \hskip 7.5 cm
    {\large IF-UERJ-#4 }

    \hskip 7.5 cm
    {\large Preprint}

    \hskip 7.5 cm
    {\large #5 }

    \medskip
    \noindent
    \hrulefill

    \vskip 2.9 cm
    }
\def\({\c c}
\def\|{\'\i}

\def\dgraphic #1#2#3#4{
    \centerline{\hbox to#1{\special{anisoscale #3, #1 #2}\hfil}} 
    \vspace*{#2} \relax         
    \begin{figure}[h] \caption{#4} \end{figure}   
    }
\def\dtwographic #1#2#3#4{
    \centerline{\hbox to#1{\special{anisoscale #3, #1 #2}\hfil}
                \hbox to#1{\special{anisoscale #4, #1 #2}\hfil}}
    \vspace*{#2} \relax
    }

\def\dthreegraphic #1#2#3#4#5{
    \centerline{\hbox to#1{\special{anisoscale #3, #1 #2}\hfil}
                \hbox to#1{\special{anisoscale #4, #1 #2}\hfil}
                \hbox to#1{\special{anisoscale #5, #1 #2}\hfil}}
    \vspace*{#2} \relax
}

\def\ni{\noindent }
\def\eq #1{Eq.(\ref{#1})}
\def\l{\left}
\def\r{\right}

\def\fr{\frac}
\def\st{$^{\rm st\,}$}
\def\ist{$^{\it st\,}$}
\def\nd{$^{\rm nd\,}$}
\def\ind{$^{\it \, nd\,}$}
\def\rd{$^{\rm rd\,}$}
\def\th{$^{\rm th\,}$}
\def\abaco#1{{\it abaco$_#1$}}
\begin{document}
\hspace\parindent
\thispagestyle{empty}

\graphic{2 in}{1.6 in}{uerj.wmf}{27/96}{July 1996}
\centerline{\LARGE \bf Computer Algebra Solving of First Order ODEs}

\bigskip
\centerline{\LARGE \bf Using Symmetry Methods}

\bigskip
\bigskip
\centerline{\large
E.S. Cheb-Terrab\footnote{Symbolic Computation Group, Departamento de F\|sica
Te\'orica, IF-UERJ.
\newline
\hspace*{.55cm}Available as http://dft.if.uerj.br/preprint/e6-27.tex},
{L.G.S. Duarte}\footnotemark[2] and {L.A.C.P. da Mota}\footnotemark[1]}

\bigskip
\bigskip
\bigskip
\begin{large}
\begin{abstract}
A set of Maple V R.3/4 computer algebra routines for the analytical solving
of 1\st order ODEs, using Lie group symmetry methods, is presented. The set
of commands includes a 1\st order ODE-solver and routines for, among other
things: the explicit determination of the coefficients of the infinitesimal
symmetry generator; the construction of the most general invariant 1\st
order ODE under given symmetries; the determination of the canonical
coordinates of the underlying invariant group; and the testing of the
returned results.

\end{abstract}

\bigskip
\centerline{ \underline{\hspace{6.5 cm}} }

\medskip
\centerline{ {\bf (Submitted to Computer Physics Communications)} }

\end{large}
\newpage
\setcounter{page}{1}

\bigskip
\bigskip
\hspace{1pc}
{\bf PROGRAM SUMMARY}
\bigskip

\begin{small}
\noindent
{\em Title of the software package:} First order ODE tools.   \\[10pt]
{\em Catalogue number:} (supplied by Elsevier)                \\[10pt]
{\em Software obtainable from:} CPC Program Library, Queen's
University of Belfast, N. Ireland (see application form in this issue)
\\[10pt]
{\em Licensing provisions:} none  \\[10pt]
{\em Operating systems under which the program has been tested:}
UNIX systems, Macintosh, DOS (AT 386, 486 and Pentium based) systems,
DEC VMS, IBM CMS.                        \\[10pt]
{\em Programming language used:} {\bf Maple V} Release 3/4. \\[10pt]
{\em Memory required to execute with typical data:}  8 Megabytes. \\[10pt]
{\em No. of lines in distributed program, including On-Line Help,
etc.:} 4183.                                                   \\[10pt]
{\em Nature of mathematical problem}\\
Analytical solving of 1\st order ordinary differential equations.\\[10pt]
{\em Methods of solution}\\
Computer algebra implementation of Lie group symmetry methods.   \\[10pt]
{\em Restrictions concerning the complexity of the problem}\\
Besides the inherent restrictions of the method (there is as yet no general
scheme for solving the associated PDE for the coefficients of the
infinitesimal symmetry generator), the present implementation does not work
with systems of ODEs nor with higher order ODEs. \\[10pt]
{\em Typical running time}\\
This depends strongly on the ODE to be solved, usually taking from a few
seconds to a few minutes. In the tests we ran (with 466 1\st order ODEs from
Kamke's book \cite{kamke}, see sec.\ref{tests}), the {\it average times}
were: 6 sec. for a solved ODE and 20 sec. for an unsolved one, using a
Pentium 133 with 64 Mb. RAM on a {\it Windows} 3.11 platform. \\[10pt]
{\em Unusual features of the program}\\
The 1\st order ODE-solver here presented is an implementation of {\it all}
the steps of the symmetry method solving scheme; i.e., when successful it
returns a closed solution, not the symmetry generator. Also, this solver
permits the user to (optionally) participate in the solving process by
giving an advice (HINT option) concerning the {\it functional form} for the
coefficients of the infinitesimal symmetry generator (infinitesimals). All
the intermediate steps of the symmetry method solving scheme are available
as user-level commands too. For instance, using the package's commands, it
is possible to obtain the infinitesimals, the related canonical coordinates,
and the most general 1\st order ODE invariant under a symmetry group. The
package also includes a command for classifying ODEs (according to Kamke's'
book\cite{kamke}) popping up Help pages with Kamke's advice for solving
them, facilitating the study of a given ODE and the use of the package with
pedagogical purposes.
\end{small}
\newpage
%
%
\hspace{1pc}
{\large {\bf LONG WRITE-UP}}

\section*{Introduction}

The automatic computation of symbolic solutions for ODEs has already been
implemented in almost all symbolic computation systems available. However,
though Lie's discovery of symmetry methods (SM) put most of the known
solving methods under a common umbrella\cite{bluman,olver}, only a few of
the computer algebra standard solvers use SM in their solving
schemes\cite{hereman}, and almost none use them for solving 1\st order ODEs.

Generally speaking, one of the reasons for this is that the use of SM
requires the solving of a system of partial differential equations (PDEs)
for the coefficients of the infinitesimal symmetry generators
(infinitesimals), whose solution is not obvious, thus involving the
programming of semi-systematic {\it heuristic} methods for obtaining
it\cite{hereman}. Also, though in the specific case of 1\st order ODEs we
just need to solve a single 1\st order PDE, one may think that SM are
useless anyway\cite{maccallum}, since to {\it guess} a solution for that PDE
would not be simpler than to {\it guess} an equivalent integrating factor
for the original ODE. Furthermore, there are many well known methods, based
on a previous {\it classification} of the ODE, which already give good
results without using heuristics\cite{kamke}. So what would be the advantage
of introducing heuristics in the solving process?

There are however two strong arguments in favor of heuristic methods: they
are the only chance of solving {\it non-classifiable} ODEs, and they are the
main way to find the underlying logic structures with which one could
develop new methods.

Bearing all this in mind, we programmed an 1\st order ODE-solver based on
SM, implementing a set of heuristic methods for solving the PDE which leads
to the infinitesimals. In addition, in order to test ODE-solvers in general,
we prepared a file with the Maple input of the 576 ODEs\footnote{From here
on, we will use the abbreviate ODE for representing 1\st order ODEs.} of
Kamke's book, and another one with 424 ODES, mainly of Riccati, Abel and
non-classifiable types.

The set of user-level routines here presented also allows the user to obtain
almost all the intermediate results involved in the SM scheme, to find the
most general invariant ODE under a given symmetry, to test explicit/implicit
results obtained by any (Maple) ODE-solver, and to classify ODEs, according
to Kamke's book, including the optional pop up of Help pages with Kamke's
{\it advice} about how to solve them.

The goals of such a work can be summarized as: \begin{enumerate} \item to
build an ODE-solver complementing the standard {\bf dsolve}, mainly for
tackling non-classifiable ODEs, \item to build a research environment for
discussing possible connections between patterns of infinitesimals
and patterns of ODEs, mainly using the HINT option of the solver (see
sec.\ref{package}) in connection with the {\it advisor} ({\bf odeadv})
command, \item to build a friendly educational environment for studying ODEs
and their solving methods in general. \end{enumerate}

The exposition is organized as follows. In sec.\ref{liemethod}, the SM
scheme for solving ODEs is briefly reviewed and discussed. In
sec.\ref{package} the package of routines is presented. Sec.\ref{algorithms}
contains a more detailed explanation of each of the heuristic
methods implemented. Sec.\ref{tests} is devoted to testing the package,
mainly the solver, against the aforementioned set of 1,000 ODEs. These
tests allow an evaluation of the solver from different angles, and include a
comparison of the performances of the new solver and the Maple {\bf dsolve}
in solving Kamke's ODEs. Finally, the conclusions contain a brief discussion
about this work and its possible extensions.

\section{Symmetry methods for 1\st order ODEs}
\label{liemethod}


Generally speaking, the key point of Lie's method for solving ODEs is that
the knowledge of a (Lie) group of transformations which leaves a given ODE
invariant may help in reducing the problem of finding its solution to
quadratures\cite{bluman,olver}. Despite the subtleties which arise when
considering different cases, we can summarize the computational task of
using SM for solving a given ODE, say,

\begin{equation}
\fr{dy}{dx}=\Phi(y,x),
\label{ode}
\end{equation}

\ni as {\it the finding of the infinitesimals} of a one-parameter Lie group
which leaves \eq{ode} invariant, i.e., a pair of functions
$\{\xi(y,x),\,\eta(y,x)\}$ satisfying

\begin{equation}
{\frac {{ \partial { \eta}}}{{ \partial}{x}}}
+  \left(  {\frac {{ \partial { \eta}}}{{ \partial}{y}}}
-  {\frac {{ \partial { \xi}}}{{ \partial}{x}}} \right) \,{ \Phi}
-  {\frac {{ \partial { \xi}}}{{ \partial}{y}}} \,{ \Phi}^{2}
- { \xi}\, {\frac {{ \partial { \Phi}}}{{ \partial}{x}}}
- { \eta}\, {\frac {{ \partial { \Phi}}}{{ \partial}{y}}} =0
\label{xi_eta}
\end{equation}

followed by either


\begin{description}
\item [a)] the determination of the canonical coordinates, say $\{r,s\}$, of
the associated Lie group, to be used in a change of variables which will
reduce \eq{ode} to a quadrature; or, alternatively,
\item[b)] the setting up of an integrating factor for \eq{ode}
using the explicit expressions found for $\{\xi(y,x),\,\eta(y,x)\}$.
\end{description}


A first look at the symmetry scheme may lead to the conclusion that finding
solutions to \eq{xi_eta} would be much more difficult than solving the
original \eq{ode}. However, what we are really looking for is a particular
solution to \eq{xi_eta}, as {\it simple as possible} (for instance two
constants), and for a non-classifiable ODE this particular solution may be
the simpler one to be obtained.

As an example, consider

\begin{equation} {\frac {{ \partial y}}{{ \partial}{x}}}= {\displaystyle
\frac {(y - x\,\ln(x))^{2}}{x^2}} + \ln(x) \label{convo1} \end{equation}

This equation is of Riccati type and cannot be solved easily\footnote{An
analysis of \eq{convo1} can be obtained by sending it in REDUCE format (to
be tackled with the CONVODE program) to
convode@riemann.physmath.fundp.ac.be}, but a polynomial {\it guess} for the
infinitesimals (made by one of {\bf symgen}'s internal routines) leads
to\footnote{In what follows, the {\it input} can be recognized by the Maple
prompt \verb->-.}:

\begin{verbatim}
> symgen(");        # input = ODE,  output = the infinitesimals
\end{verbatim}
\begin{displaymath}
\_\xi(x,y)=x,\ \ \ \ \_\eta(x,y)=x+y
\end{displaymath}

Passing \eq{convo1} directly to {\bf odsolve} (the solver), it will call
{\bf symgen}, use the result above to determine the canonical coordinates,
change the variables, and solve \eq{convo1} as follows:

\begin{verbatim}
> odsolve("");
\end{verbatim}
\begin{displaymath}
y={\frac {x\sqrt {5}}{2}}\
{\tanh\l(-{\frac {\sqrt {5}\ln (x)}{2}}+{\frac {
\sqrt {5}{\it \_C1}}{2}}\r)}
+x\,\ln (x)+{\frac {x}{2}}
\end{displaymath}

What is amazing, and characteristic of symmetry methods is that changing $(y
- x\,\ln(x))/x$ to $F((y - x\,\ln(x))/x)$ in \eq{convo1}, where $F$ is an
arbitrary function, the symmetry remains unchanged and the solving scheme
succeeds as well:

\begin{equation}
{\frac {dy}{dx}}=F\l({\frac {y-x\,\ln (x)}{x}}\r)+\ln (x)
\label{intat1}
\end{equation}
\begin{verbatim}
> odsolve(");
\end{verbatim}
\begin{displaymath}
\ln (x)=-{\int}^{\displaystyle \frac {y-x\,\ln (x)}{x}}
\fr {d\_a} {(1+{\it \_a}-F({\it \_a}))}
+{\it \_C1}
\end{displaymath}

The integral above is expressed using the new {\bf intat} command (of the
new version of {\it PDEtools} \cite{PDEtools}) which represents the {\it
integral evaluated at a point} -analogous to the derivative evaluated at a
point. {\bf intat} displays the {\it evaluation point} as an upper limit of
integration. For \eq{intat1}, Maple V R.4 returned an {\it explicit} result
actually equivalent to the above (see \eq{intat2}).

Even when the built-in heuristics fail, in the framework of SM, it is
possible to provide computational routines permitting users to test their
own conjectures concerning the {\it functional form} of the infinitesimals
(HINT option, see sec.\ref{package}). As an example of this, consider

\begin{equation}
{\frac {dy}{dx}}={\frac {x+\cos({{\rm e}^{ y}}+\left( x+1\right)
{\rm e^{ -x}})}{{{\rm e}^{(y+x)}}}}
\label{convo2}
\end{equation}

This ODE is not classifiable nor is it solved by the automatic heuristics of
{\bf odsolve}. An example of a user-conjecture concerning the infinitesimals
would be:

\begin{equation}
\xi(x,y)=f(x),\ \ \ \ \eta(x,y)=x\,g(y)
\end{equation}

where $f(x)$ and $g(y)$ are unknown functions of their arguments. The {\tt
HINT} algorithm we programmed, given such an ansatz, will try to determine
$f$ and $g$, in this case leading to:

\begin{verbatim}
> symgen(",HINT=[f(x),x*g(y)]);
\end{verbatim}
\begin{displaymath}
{\_ \xi}=e^{{x}},\ \ \ \ {\_ \eta}={x}\,e^{-y}
\end{displaymath}
\begin{verbatim}
> odsolve("",HINT=[f(x),x*g(y)]);
\end{verbatim}
\begin{displaymath}
y=\ln \left(2\,\arctan\l({\frac
{{e^{-\left ({e^{-x}}+{\it \_C1}\right )}}-1}
{{e^{-\left ({e^{-x}}+{\it \_C1}\right )}}+1}
}\r)-(x+1)\,{e^{-x}}\r)
\end{displaymath}

\section{The package's commands}
\label{package}

A detailed description of the package's commands, with examples and
explanations concerning their calling sequences, is found in the On-Line
Help. Therefore, we have restricted this section to a brief summary,
followed by a detailed description of the solver, {\bf odsolve}, and the
advisor, {\bf odeadv}, and a description of just one paragraph for each of
the other routines\footnote{This section contains some information already
presented in the previous section; this was necessary to produce a complete
description of the package.}. Simple {\it input/output} examples can be seen
in sec.\ref{algorithms}.

\subsection{\it Summary}
A brief review of the commands of the package is as follows:

\bigskip
{\centerline
{\small
\begin{tabular}{|p{1.2 in} |l|}
\hline
Command 	  & Purpose: \\
\hline
{\bf odsolve} & solve ODEs using the symmetry method scheme \\
{\bf fatint } & look for an integrating factor\\
{\bf canoni } & look for a pair of canonical coordinates\\
{\bf symgen}  & look for a pair of infinitesimals\\
{\bf equinv}  & look for the most general ODE invariant under a given 
symmetry \\
{\bf buisym } & look for the infinitesimals given the solution of an ODE\\
{\bf odepde}  & return the PDE for the infinitesimals\\
{\bf odetest} & test explicit/implicit results obtained by any Maple 
ODE-solver \\
{\bf symtest} & test a given symmetry\\
{\bf odeadv}  & classify ODEs and pop up a Help-page with Kamke's
solving advice\\
\hline
\end{tabular} }}

\subsection{\it Description}
\label{description}

\subsubsection*{Command name: {\bf odsolve}}
\label{odsolve}

\noindent {\it Feature:} 1\st order ODE-solver based on symmetry methods

\noindent {\it Calling sequence:}
\begin{verbatim}
> odsolve(ode);}
> odsolve(ode, y(x), way=xxx, HINT=[expr1, expr2], int_scheme);}
\end{verbatim}

\noindent
{\it Parameters:}

\bigskip
\noindent
\begin{tabular}{p{1.2in}p{4.9in}}
\verb-ode- & - a 1\st order ODE. \\
\verb-y(x)- & - the dependent variable (required when not obvious).\\
\verb-way=xxx- & - optional, forces the use of only one ({\tt xxx}) of the 6
internal algorithms \{abaco1, 2, 3, 4, 5, abaco2\} for determining the
coefficients of the infinitesimal symmetry generator (infinitesimals).\\
\verb-HINT = [e1,e2] - & - optional, {\tt e1} and {\tt e2} indicate a
possible {\it functional} form for the infinitesimals. \\
\verb-int_scheme- & - optional, one of:
\verb-fat-, \verb-can-; meaning respectively:
use an integrating factor; use canonical
coordinates.
\end{tabular}

\noindent
Optional parameters can be given alone or in conjunction, and in any order.

\noindent
{\it Synopsis:}

\medskip Given a 1\st order ODE, {\bf odsolve}'s main goal is to solve it in
two steps: first, determine the infinitesimals of a 1-parameter symmetry
group which leaves the ODE invariant, and then use these infinitesimals to
integrate the ODE, by building an integrating factor or reducing the ODE to
a quadrature using the canonical coordinates of that group. To determine the
infinitesimals, {\bf odsolve} makes calls to {\bf symgen}, another command
of the package, and to make the change of variables introducing the
canonical coordinates, it makes calls to {\bf dchange}, from the {\it
PDEtools} package \cite{PDEtools}.

{\bf odsolve} does not {\it classify} the ODE before tackling it, and is
mainly concerned with {\it non-classifiable} ODEs, for which the standard
{\bf dsolve} fails. By default, {\bf odsolve} starts off trying to isolate
the derivative in the given ODE, then sequentially uses all six internal
algorithms of {\bf symgen} to try to determine the infinitesimals (see
sec.\ref{algorithms}), and finally tries the two integration schemes
sequentially.


When {\bf odsolve} succeeds in solving the ODE, it returns, in order of
preference: an {\it explicit} closed solution, an {\it implicit} closed
solution, or a {\it parametric} solution (only when the derivative cannot be
isolated\cite{BRA}).

All these defaults can be changed by the user. The main options she/he has
are:

\begin{enumerate}

\item to request the use of only one of the six internal algorithms for
determining the infinitesimals ({\tt way=xxx} option; different algorithms
might lead to different symmetries for the same ODE, perhaps turning the
integration step easier);

\item to enforce the use of only one of the two alternative schemes for
integrating a given ODE after finding the infinitesimals ({\tt can} or {\tt
fat} optional arguments, useful to select the best integration strategy for
each case);

\item to indicate a possible {\it functional form} for the infinitesimals
({\tt HINT=xxx} option). This option is valuable when the solver fails, or
to study the possible connection between the algebraic pattern of the given
ODE and that of the infinitesimals.

\item to enforce the use of {\bf dsolve} at first and {\bf symgen}'s
algorithms only when it fails (assign the global variable {\tt sym :=
false});

\end{enumerate}

A brief description of how the {\tt HINT=xxx} option can be used (see
examples in sec.\ref{liemethod} and \ref{algorithms}) is as follows:

\begin{itemize}

\item {\tt HINT=[e1,e2]} indicates to the solver that it should take e1 and
e2 as the infinitesimals and determine the form of any (a maximum of two)
indeterminate functions entering e1 and/or e2, such as to solve the
problem (e.g., \eq{convo2}).

\item {\tt HINT=[e1,`*`]} where e1 represents the user's guess for the first
infinitesimal, $\xi$, and `*` indicates to the solver that it must consider
the other infinitesimal, $\eta$, as a product of two indeterminate functions
of one variable each: the independent and the dependent variables,
respectively (e.g., \eq{Bernoulli}).

\item {\tt HINT=[e1,`+`]} works as in the previous case, but replacing the
product of two indeterminate functions by a sum of them .

\item {\tt HINT=parametric} indicates to the solver that it should look for
a parametric solution for the given ODE. This solving scheme may be useful
even when the system succeeds in isolating the derivative \cite{BRA}.
\end{itemize}

Finally, there are three global variables managing the solving process,
which are automatically set by internal routines but can also be set by the
user, as desired. They are \{{\it dgun, ngun, sgun}\}, for setting,
respectively, the maximum {\it d}egree of some polynomials entering
ans\"atze for the infinitesimals, the maximum {\it n}umber of subproblems
into which the original ODE should be mapped, and the maximum {\it s}ize
permitted for such subproblems. The {\it dgun} variable is automatically set
each time {\bf symgen} is called, according to the given ODE, whereas, by
default, {\it ngun} and {\it sgun} have their values respectively fixed to
$2$ and $1$ . Although the setting of these variables by the user might
increase the efficacy of the algorithms significantly, increasing by one
unit the {\it ngun} or {\it sgun} variables, depending on the given ODE, may
geometrically slow down the solving process.

Apart from the solver, the package includes an analyzer, {\bf odeadv}, which
can play a pedagogical role, or be a comfortable tool for classifying ODEs
while studying the possible links between symmetry methods and standard
ones. A detailed description of it is as follows:

\medskip
\noindent
\subsubsection*{Command name: {\bf odeadv}}
\label{odeadv}

\medskip
\noindent
{\it Feature:} 1\st order ODE-analyzer and solving advisor

\medskip \noindent {\it Calling sequence:}
\begin{verbatim}
> odeadv(ode);
> odeadv(ode, y(x),{type1,type2,type3,...},help);
\end{verbatim}

\medskip
\noindent
{\it Parameters:}

\medskip
\noindent
\begin{tabular}{p{1.4in}p{4.7in}}
\verb-ode- & - a 1\st order ODE \\
\verb-y(x)- & - the indeterminate function (necessary when not obvious)\\
\verb-{type1,type2,...}- & - optional, a subset of ODE classification types
to be checked \\
\verb-help- & - optional, to request the pop-up of a
Help-page indicating Kamke's book advice for solving the given ODE.
\end{tabular}

\noindent

\medskip
\noindent
{\it Synopsis:}

\medskip Given a 1\st order ODE, {\bf odeadv}'s main goal is to {\it
classify} it according to Kamke's book and pop up a Help-page indicating
Kamke's advice for solving it when required (by entering the word
\verb-help- as an extra argument). The Help-pages include examples and the
Maple input lines implementing the advice, so as to allow the user to adapt
them to her/his problem.

When used without extra arguments, {\bf odeadv} will try to classify the
given ODE into one or more of the following types:

\begin{quote} {\it quadrature; separable; linear: classes A, B, and C;
homogeneous: classes A, C and D; exact; rational; Clairaut; Bernoulli;
Riccati, Riccati special; Abel: 1\ist type, 2\ind type subclasses A, B and
C; d'Alembert; and ``patterns": $y=g(y')$, $x=g(y')$, $0=G(x,y')$,
$0=G(y,y')$, $y=G(x,y')$, $x=G(y,y')$} \end{quote}


The matching of the types is checked sequentially, in the order displayed
above. {\bf odeadv} may return more than one type when the ODE is of type
homogeneous, exact or rational; otherwise, the first matching of a pattern
will interrupt the process and a classification will be returned.

As an option, the user can indicate the checking of just a subset of the
types mentioned above, by giving this subset as an extra argument.

A compact description of the purpose of the other routines of the package is
as follows:

\begin{itemize}

\item {\bf symgen} looks for the infinitesimals $\xi(x,y)$ and $\eta(x,y)$
of a 1-parameter Lie group which leave the received ODE invariant, as a pair
of functions satisfying \eq{xi_eta}. The options available for this command
are the same {\tt way=xxx} and {\tt HINT=xxx} options of {\bf odsolve}.

\item {\bf fatint} and {\bf canoni} respectively return the integrating
factor for the given ODE or a set of transformations from the original
coordinates to the canonical coordinates of the Lie group mentioned above.

\item {\bf equinv} takes as arguments a list of two algebraic expressions,
to be seen as the infinitesimals of a 1-parameter Lie group, and returns,
within the possibilities of the system, the most general ODE invariant under
that group.

\item {\bf buisym} takes as arguments the solution of an ODE and an
indication of which are the dependent and independent variables, and looks
for a pair of infinitesimals for the ODE which generated the problem.

\item {\bf odetest} tests explicit and implicit solutions found by {\it any}
Maple ODE-solver, such as {\bf dsolve} or {\bf odsolve}, by making a careful
simplification of the ODE with respect to that solution.

\item {\bf symtest} tests the results returned by {\bf symgen}, returning 0
when these results check {\it ok}, or an algebraic expression obtained after
simplifying the PDE for the infinitesimals according to {\bf symgen}'s
result.

\end{itemize}

\noindent
{\bf buisym} and {\bf equinv} may be useful in connection with the {\bf
odeadv} command and the {\tt HINT} option of {\bf odsolve}, in order to
study the relation between symmetry patterns and ODE patterns. Also, it is
worth mentioning that, since any {\it solver} will make extensive use of the
whole Maple library, testing commands such as {\bf odetest} and {\bf
symtest} would also be useful in detecting possible wrong results of the
Maple library.

\section{The {\it heuristic} methods}
\label{algorithms}

\subsection{General remarks}

All the implemented {\it heuristic} methods assume that the system was
successful in isolating the derivative, as in \eq{ode}. These methods
consist of six computational {\it algorithms} for building and testing an
explicit algebraic expression for the infinitesimals $\xi$ and $\eta$. By
heuristic we mean that the possible success of the scheme cannot be
determined {\it a priori}.

Two of the six algorithms are in turn subdivided into eight and four
subalgorithms respectively; thus, in all, the given ODE is actually tackled
by trying sixteen different {\it functional forms} for $\xi$ and $\eta$.
Three of these sixteen schemes involve different types of polynomial
ans\"atze for $\{ \xi,\ \eta \}$, whose form, apart from the polynomial
degree, is virtually independent of the ODE input. In the other thirteen
schemes, the ans\"atze for $\xi$ and $\eta$ depend on the received problem,
and are not {\it a priori} of polynomial type, but rather are obtained by
solving auxiliary ODEs.

Four of the six algorithms directly tackle \eq{xi_eta}, while the other two
reformulate the problem taking into account that \eq{xi_eta} admits
$\eta=\xi\,\Phi$ as a general (useless) solution. Therefore, it is always
possible to introduce

\begin{equation}
\eta=\xi\,\Phi+\chi
\label{constraint}
\end{equation}

in \eq{xi_eta}, where $\chi(x,y)$ is an unknown function of its arguments,
in order to map the original problem into that of solving

\begin{equation}
{\frac {\partial \chi}{\partial x}}
+\Phi{\frac {\partial \chi}{\partial y}}
-\left ({\frac {\partial \Phi}{\partial y}} \right )\chi=0
\label{chi}
\end{equation}

for $\chi$. The knowledge of $\chi$, in turn, allows one to set $\xi$ and
$\eta$ as desired using \eq{constraint}.

For the case in which all these schemes fail, we programmed a HINT option to
allow the user to try her/his own heuristics for finding the infinitesimals,
as explained in sec.\ref{odsolve}.

\subsection*{The six algorithms}

The 1\st algorithm, called \abaco{1}, consists of 2 sets of 4 schemes each,
leading to 8 different trials. The first 4 trials look for $\{ \xi,\ \eta
\}$ taking one of the infinitesimals as 0 and the other one as a function of
only one variable, namely:

\begin{equation}
\{ \xi=0,\,\eta=f(x) \},\ \ \
\{ \xi=0,\,\eta=f(y) \},\ \ \
\{ \xi=f(x),\,\eta=0 \},\ \ \
\{ \xi=f(y),\,\eta=0 \}
\end{equation}

The possible success of each case is determined by algebraic factorization
of \eq{xi_eta}, and when this possibility is confirmed, an auxiliary ODE is
built to determine the explicit resulting form for $f$.

The second sequence of 4 trials looks for $\{ \xi,\ \eta \}$ taking one of
the infinitesimals as 0 and the other one as an expression containing both $x$
and $y$, namely:

\begin{eqnarray}
\{ \xi=0,\,\eta=f(x)\,g(y) \}, & \ \ \
\{ \xi=f(x)\,g(y),\,\eta=0 \},\nonumber \\*[.1in]
\{ \xi=0,\,\eta=f(y)\,g(x) \}, & \ \ \
\{ \xi= f(y)\,g(x),\,\eta=0 \}
\label{abacomagico}
\end{eqnarray}

where $g$ is an algebraic expression built by extracting factors from the
numerator and denominator of the ODE to be solved, and $f$ is an unknown
function to be determined by solving auxiliary ODEs as in the first 4
trials.

\noindent {\it Example\footnote{In all the following examples, an alias has
been set for $y$, via \verb-> alias(y=y(x));-}:}  Kamke's ODE 120.

\begin{equation}
x{\frac {dy}{dx}}-y\l(x\ln \l({\frac {{x}^{2}}{y}}\r)+2\r)=0
\end{equation}
\begin{verbatim}
> symgen(");
\end{verbatim}
\begin{displaymath}
{\it \_\xi}=0,\ \ \ \ {\it \_\eta}=-{e^{-x}}y
\ \ \ \ \ \ \ \ \ \ \
\mbox{($g(y)=y$ and $f(x)$ was determined as $-e^{-x}$)}
\end{displaymath}
\begin{verbatim}
> odsolve("");
\end{verbatim}
\begin{displaymath}
y={x}^{2}{e^{({\it \_C1}\,{e^{-x}})}}
\end{displaymath}

As a general benchmark, \abaco{1} can determine a pair of infinitesimals for
151 of the 466 non-quadratures found in Kamke's first 500 ODEs, spending on
each an average time of 0.26 sec. when successful and 1.1 sec. otherwise.

The 2\nd of the set of six main algorithms consists of a bivariate
polynomial in $(x,y)$ ansatz for $\chi$ to solve \eq{chi}. The maximum
degree of the polynomial can be set by the user by assigning the global {\it
dgun} variable; otherwise, it will be determined automatically by an
internal routine. This second algorithm can determine a pair of
infinitesimals for 96 of Kamke's 466 non-quadrature ODEs mentioned above,
spending an average time of 0.08 sec. when successful and 0.5 sec.
otherwise.

The 3\rd and 4\th of the set of six algorithms respectively consist of
bivariate polynomial and rational ans\"atze in $(x,y)$ for $\{\xi,\,\eta\}$,
to solve \eq{xi_eta}.

\noindent {\it Example:} Kamke's ODE 236 (Abel, 2\nd type, class B),

\begin{equation}
x\left(y+4\right)\,{\frac {dy}{dx}}-y^{2}-2\,y-2\,x=0
\end{equation}
\begin{verbatim}
> symgen(");
\end{verbatim}
\begin{displaymath}
{\it \_\xi}=4\,x+{x}^{2},\ \ \ \ {\it \_\eta}=xy+4\,x
\end{displaymath}

In solving the aforementioned 466 Kamke's non-quadrature ODEs, these two
algorithms can determine a pair of infinitesimals in 234 cases, spending an
average time of 0.7 sec. in each success and 1.2 sec otherwise.

The 5\th algorithm builds an explicit expression for $\chi(x,y)$, to solve
\eq{chi}, as follows:

\begin{enumerate}

\item a basis of functions and algebraic objects is built by taking, from
the given ODE, all the {\it known} functions\footnote{By {\it known}
function we mean a function whose derivative rule is known by the Maple
system.} and composite algebraic objects, together with their derivatives,
as well as all the {\it unknown} functions;

\item a polynomial of degree 2 in such objects is built; its coefficients,
in turn, are polynomials of degree $d$ (the {\it dgun} variable mentioned
above) in $\{x,y\}$ with undetermined coefficients.

\end{enumerate}

This ansatz for $\chi$ is introduced in \eq{chi} resulting in a system of
algebraic equations for the undetermined coefficients mentioned above.

\noindent {\it Example:} Kamke's ODE 357 (no classification)

\begin{equation}
x\left ({\frac {dy}{dx}}\right )\ln (x)\sin(y)+\cos(y)\left (1-
x\cos(y)\right )=0
\end{equation}
\begin{verbatim}
> symgen(");
\end{verbatim}
\begin{displaymath}
{\it \_\xi}=0,\ \ \ \
{\it \_\eta}=\l( {{\frac {\cos(2\,y)x}{2}}+{\frac {x}{2}}}\r)\,
{\frac {1}{2\,x\ln (x)\sin(y)}}
\end{displaymath}
\begin{verbatim}
> odsolve("");
\end{verbatim}
\begin{displaymath}
y=\arccos\l({\frac {2\,\ln (x)}{2\,x+{\it \_C1}}}\r)
\end{displaymath}

This algorithm can find a pair of infinitesimals for 218 of Kamke's 466
non-quadratures mentioned above, spending on each an average time of 2.1
sec. when successful and 1.8 sec. otherwise.

The 6\th algorithm, called \abaco{2}, leads to a sequence of 4 trials for
the infinitesimals $\{\xi,\,\eta\}$, making use of the HINT option of {\bf
symgen}, with two indeterminate functions of only one variable each, namely:

\begin{eqnarray}
\{ \xi=g(x),\,\eta=f(x) \}, & \ \ \
\{ \xi=g(x),\,\eta=f(y) \},\nonumber \\*[.07in]
\{ \xi=g(y),\,\eta=f(x) \}, & \ \ \
\{ \xi=g(y),\,\eta=f(y) \}
\label{abaco2}
\end{eqnarray}

The possible success of these trials {\it cannot} be determined by algebraic
factorization of \eq{xi_eta}, as in \abaco{1}, except in a few cases. Hence,
the alternative scheme we implemented for managing such HINTs consists of
determining $f$ and $g$ as follows:
\begin{enumerate}
\item subdivide \eq{xi_eta} into subexpressions involving only one of 
$\{f,\,g\}$;
\item build a list of {\it candidates} for $f$ and for $g$ with the
solutions to these subexpressions;\label{listfg}
\item build a list of {\it
pairs of candidates} by taking one candidate from each list.
\end{enumerate}

To avoid spending much time, the number of {\it candidates} of each list in
step (\ref{listfg}) is restricted, by default, to $n=2$, leading to a
maximum of 4 {\it pairs of candidates} for each functional form in
\eq{abaco2}. This default can be changed by the user by assigning
the global variable {\it ngun}.

\noindent {\it Examples:} ODE 593 (file 6) with an arbitrary function $F(x,y)$

\begin{equation}
{\frac {dy}{dx}}={\frac {{e^{x}}} {\sqrt {y}}}\,{F\l(\sqrt{y^3}-
{\frac {3\,{e^{x}}}{2}}\r)}
\label{intat2}
\end{equation}
\begin{verbatim}
> symgen(");
\end{verbatim}
\begin{displaymath}
{\it \_\xi}={\frac {1}{{e^{x}}}},\ \ \ \
{\it \_\eta}={\frac {1}{\sqrt {y}}}
\ \ \ \ \ \ \ \ \ \ \
\mbox{($g(y)=\displaystyle \fr{1}{\sqrt{y}}$, $f(x)=\displaystyle
\fr{1}{e^x}$ - see \eq{abaco2})}
\end{displaymath}
\begin{verbatim}
> odsolve("");          # Maple V R.4 output using intat and RootOf
\end{verbatim}
\begin{displaymath}
y=\left ({\rm RootOf}(
{\int}^{
\fr{2\,{{\it \_Z}}^{3}}{3}-{e^{x}}}
\fr {{\it d\_a}} {(-1+F(\fr{3\,\_a}{2}))}-{e^{x}}+{\it
\_C1})\right )^{2}
\end{displaymath}
\begin{verbatim}
> odetest(""","); # explicit and implicit results can be tested with odetest
\end{verbatim}
\begin{displaymath}
0
\end{displaymath}

For this ODE, Maple V R.3 returned an {\it implicit} (equivalent) result, as
in \eq{intat1}.

\abaco{2} can determine $\{\xi,\,\eta\}$ for 114 of Kamke's 466
non-quadratures mentioned above, spending on each an average time of 0.4
sec. when successful and 8 sec. otherwise.

To conclude this section, a curious result illustrating the HINT option is
given by the infinitesimals found by {\bf symgen} for the general form of
Bernoulli's equation,

\begin{equation}
{\frac {dy}{dx}}+f(x)\,y+g(x)\,y^{\alpha}=0,
\label{Bernoulli}
\end{equation}

where $f$ and $g$ are arbitrary functions of their arguments, and $\alpha$
is a symbolic power. The giving of a HINT, asking for a {\it separation of
variables by product} for one of the infinitesimals (see
subsec.\ref{odsolve}) leads to:

\begin{verbatim}
> symgen(",HINT=[0,`*`]);
\end{verbatim}
\begin{displaymath}
{\it \_\xi}=0,\ \ \ \
{\it \_\eta}={y}^{\alpha}\,
{e^{\displaystyle \int\! f(x){dx}\,(\alpha-1)}}
\end{displaymath}
\begin{verbatim}
> factor(odsolve("",HINT=[0,`*`]));
\end{verbatim}
\begin{displaymath}
y=\left({e^{\displaystyle \l({\int \!f(x){dx}\left (\alpha-1\right )}\r)}}
\left (\alpha-1\right)\left ({\int}^{x}g({\it \_a})
{e^{\displaystyle \l({-\int \!f({\it \_a}){d{\it \_a}}
\left (\alpha-1\right)}\r)}}{\it d\_a}-{\it \_C1}
\right )\right )^{\frac{1}{1-\alpha}}
\end{displaymath}

\section{Tests}
\label{tests}

We tested the set of routines here presented having two different ideas in
mind: to confirm the correctness of the returned results, and to evaluate
the new routines' {\it performance} in solving Kamke's ODEs. The performance
test was done with the intention of comparing the efficacy of a {\it
SM-based} solver such as {\bf odsolve} with that of a {\it
classification-based} solver as is Maple's {\bf dsolve}.

We used 10 test files, containing a set of 1,000 ODEs distributed in: files
1 to 6 containing the 576 ODEs of Kamke's book, and files 7 to 10 containing
424 ODEs of Riccati, Abel and {\it non-classifiable} types, for which the
{\it classification-based} {\bf dsolve} (R.3/4) fails. The ODEs of files
7-10 were built using the {\bf equinv} command, departing from symmetries of
Kamke's ODEs, and were used to test the correctness of the results returned
by {\bf symgen} and the integration subroutines of {\bf odsolve}. The ``{\it
fail} by {\bf dsolve}" condition for these 424 ODEs was proposed to evaluate
{\bf odsolve} in solving the type of ODEs for which it was really designed,
i.e., non-classifiable or usually not solved by {\bf dsolve}.

The purpose of preparing files with such a big set of ODEs was to set up
what would be a convenient test-file for evaluating the performance of
computational implementations in solving {\it classifiable}/{\it
non-classifiable} ODEs. The 10 files, with the input code in Maple, Reduce
and Mathematica format, are available at our site:
http://dft.if.uerj.br/symbcomp.html.

A summary with the number of ODEs of each file belonging to each {\it
classification} (we used our {\bf odeadv} classification command) is as
follows:

{\begin{center} {\footnotesize
\begin{tabular}{|l|c|c|c|c|c|c|c|c|c|c|c|}
\hline
 & \multicolumn{10}{|c|}{File:} & \\
\cline{2-11}
Class:  &  1 &2  &  3 &  4 &  5 &  6 &  7 &  8 &  9 & 10 & Totals:\\
\hline
quadrature  & 10 &  0 &  1 & 12 & 10 & 13 &  0 &  0 &  0 &  0 & 46 \\
\hline
separable   & 19 & 10 &  3 &  8 & 10 &  7 &  0 &  0 &  0 &  0 & 57 \\
\hline
linear      & 14 & 15 & 16 &  1 &  1 &  0 &  0 &  0 &  0 &  0 & 47 \\
\hline
homogeneous & 1  & 19 & 15 & 11 & 28 &  4 & 14 & 19 & 19 &  8 & 138 \\
\hline
exact       & 0  &  0 & 14 & 14 &  0 &  0 &  0 &  0 &  0 &  0 & 28 \\
\hline
rational    & 5  & 48 & 68 & 28 & 26 &  1 & 17 & 64 & 54 & 40 & 351 \\
\hline
Clairaut    & 0  &  0 &  0 &  7 & 12 &  8 &  0 &  0 &  0 &  0 &  27\\
\hline
Bernoulli   & 0  &  1 &  0 &  0 &  0 &  0 &  0 &  0 &  0 &  0 & 1 \\
\hline
Riccati     & 23 & 46 &  1 &  0 &  0 &  0 & 13 & 27 & 19 & 21 & 150\\
\hline
Abel        & 15 &  8 & 30 &  0 &  0 &  0 & 12 & 19 & 25 & 42 & 151 \\
\hline
d'Alembert  & 3  & 2 & 15 &  7 &  36 &  10 &  1 &  0 &  0 &  0 & 74 \\
\hline
none        & 16 & 11 &  6 & 27 & 23 & 36 & 91 & 27 & 25 & 42 & 304 \\
\hline
has $F(x,y)$ &21 &  4 &  8 &  8 &  1 & 13 & 43 &  0 &  0 &  7 & 105 \\
\hline
\hline
Total of ODEs &100 & 100 & 100 & 100 & 100 & 76 & 124 & 100 & 100 & 100 & 
1,000\\
\hline
\multicolumn{12}{c}{Table 1. \it Classification of the 1,000 ODEs of 
files 1-10}\\
\end{tabular}}
\end{center}}

The quadratures pointed out in the table were found after isolating the
derivative, receiving a rhs not containing $x$ or $y(x)$ (the independent
and dependent variables all along the 1,000 ODEs). The next-to-last line of
the table above indicates the number of ODEs having an arbitrary function
$F(x,y)$; e.g. $F(x^2-e^y/y)$. Also, in many cases, {\bf odeadv} returned
more than one classification for a given ODE; all these possible
classifications are reflected above.

\subsection{Test of {\bf symgen}}

As explained in sec.\ref{liemethod}, the SM approach for solving ODEs
involves two main steps: the determination of a pair of infinitesimals and
their subsequent use in the integration process. We therefore divided the
tests into the same two steps.

The first {\it performance} test, thus, concerned the explicit
determination, by the {\bf symgen} command, of infinitesimals for Kamke's
ODEs (only the 466 non-quadratures of the first 500 ODEs were used). The
table below displays the total number of successes, the average time spent
 with each solved/fail case, and the number of successes of {\it each}
of {\bf symgen}'s six algorithms when the whole set of ODEs was tackled
using only one of them. The second column of the table indicates the number
of non-quadrature ODEs per file.
{\begin{center} {\footnotesize
\begin{tabular}
{|c|c|c|c|c|c|c|c|c|c|c|}
\hline
 & & & \multicolumn{2}{|c|}{Average time\footnotemark} &
\multicolumn{6}{|c|}{$\xi$ and $\eta$ can be determined by} \\
\cline{4-11}
File & ODEs & Successes & {\it ok} & {\it fail} & \abaco{1} & 2 & 3 & 4 & 5
& \abaco{2} \\
\hline
1 &  90 & 44 & 1.7 sec. & 4.2 sec. & 33 & 13 & 21 & 25 & 29 & 23 \\
\hline
2 & 100 & 71 & 1.2 sec. & 2.0 sec & 30 & 13 & 47 & 46 & 39 & 27 \\
\hline
3 &  99 & 81 & 0.7 sec. & 3.6 sec. & 35 & 36 & 55 & 46 & 40 & 15 \\
\hline
4 &  88 & 72 & 3.8 sec. & 4.1 sec. & 38 & 28 & 44 & 45 & 49 & 21 \\
\hline
5 &  89 & 75 & 1.7 sec. & 5.1 sec. & 15 & 6 & 67 & 71 & 61 & 28 \\
\hline
\hline
Totals: & 466 & 343 & $ \approx 8$ sec. & $\approx 12$ sec. & 151 & 96 & 234
& 233 & 218 & 114\\
\hline
\multicolumn{11}{c}
{Table 2. \it Kamke's ODEs for which the infinitesimals were determined by 
{\bf symgen}: $73 \%$}\\
\end{tabular} }
\end{center}}

\footnotetext{Due to the large number of ODEs being tested, we interrupted
and computed as {\it fail} all calculations consuming more than a few
minutes. This was the case in Kamke's ODEs numbers 51, 340, 347, 373, 394,
460 and 479; they were recalculated separately, receiving an {\it answer} in
$5 \sim 20 $ min. for all but ODE 479 .}

\subsection{Test of {\bf odsolve}'s integration schemes and comparison of
performances}
\label{comparison}

The two integration alternatives we programmed for the second step of the
solving scheme consist of either using an integration factor or making a
change of variables (canonical coordinates) reducing the ODE to a
quadrature. Of course, the corresponding routines work {\it only if} {\bf
symgen} succeeds in determining the infinitesimals.

Also, as mentioned, though {\bf odsolve} was thought as a complement to the
Maple {\bf dsolve}, we prepared a comparative test of performances to have
an idea of the possible efficacy of a {\it SM-based} solver, if compared to
that of a {\it classification-based} solver as {\bf dsolve}. With this
purpose, we tested the standard {\bf dsolve} with the non-quadrature Kamke's
ODEs too. The results we found in testing the integration schemes and in
comparing {\bf odsolve}'s performance with that of {\bf dsolve} can be
summarized as follows:
{\begin{center}
{\footnotesize
\begin{tabular}{|c|c|c|c|c|c|}
\hline
& & \multicolumn{2}{|c|}{Solved by} & \multicolumn{2}{|c|}
{Successful integration via}\\
\cline{3-6}
File & ODEs & {\bf dsolve} & {\bf odsolve}/{\bf symgen}  & {\tt can}
& {\tt fat} \\
\hline
 1 &  90 & 47 & $44\,/\,44$ & 41 &  40  \\
\hline
 2 & 100 & 71 &  $71\,/\,71$ & 66 & 70 \\
\hline
 3 &  99 & 75 &  $79\,/\,81$ & 61 & 78 \\
\hline
 4 &  88 & 63 &  $70\,/\,72$ & 51 & 68 \\
\hline
 5 &  89 & 69 &  $70\,/\,75$ & 69 & 55 \\
\hline
Totals: & 466 & 325& $334\,/\,343$ & 288 & 311 \\
\hline
\multicolumn{6}{c}
{Table 4. \it Comparative performance}\\
\end{tabular} }
\end{center}}

It was a surprise to us that, even when more than $80\, \%$ of the ODEs used
in the test are of {\it classifiable} type (see Table 1.), the performance
of {\bf odsolve}, which does not classify the received ODE before tackling
it, was almost the same as that of {\bf dsolve}\footnote{The possible
success in solving Kamke's ODEs by introducing simple {\it hypotheses} for
the infinitesimals has already been pointed out in a previous work by B.
Char \cite{char}.}. Also, {\bf odsolve}'s integration schemes succeeded in
$97\, \%$ (334 of 343) of the cases in which {\bf symgen} determined a pair
of infinitesimals, convincing us in that the main problem is, actually, the
determination of those infinitesimals, not their posterior use in the
integration process.

\section{Conclusions}

This paper presented a computer algebra implementation of symmetry methods
for solving 1\st order ODEs. Despite the additional complications introduced
by the necessity of solving an auxiliary 1\st order PDE, this method proved
to be a valuable tool for solving non-classifiable (or even classifiable)
ODEs, as shown in sec.\ref{tests}, resulting in the extension to Maple's
{\bf dsolve} we were looking for.

Moreover, we would like to remark the {\it interactive} character of the
solver too; i.e., the user is given the possibility of participating in the
solving process (the HINT option, see sec.\ref{description}), achieving in
this way a significant extension of the solving capabilities of the scheme.

Also, the computational routines presented here were designed not only for
solving ODEs: the {\bf odeadv} command and the HINT option of {\bf odsolve}
were thought as a combination for investigating new solving methods, and the
set of commands here presented was intended to be complete from the
educational point of view. Actually, with them it is possible to tackle an
ODE in the framework of symmetry methods, directly or step by step, it being
also possible to {\it go back} using the {\bf equinv} and {\bf buisym}
commands, completing the conceptual cycle.

On the other hand, this is a first version of the package and many things
can be improved. To mention but a few, the classification abilities of {\bf
odeadv} can be extended, and used by {\bf symgen} in order to determine the
most efficient ordering of the algorithms for tackling the given ODE (at
present this order is unmovable); also, a study of the fail examples can be
used as a basis for building new algorithms, mainly for ODEs of Riccati and
Abel types. Such new algorithms can be implemented straightforwardly using
the {\it HINT} option of {\bf symgen}. Moreover, for instance, it is
possible to extend the ideas presented here to tackle higher order ODEs too.
These subjects are part of our present interests and we expect to be able to
report related work in the near future\cite{BRA}.

\section*{Acknowledgments}

\noindent This work was supported by the State University of Rio de Janeiro
(UERJ), Brazil. The authors would like to thank J.E.F. Skea\footnote{Symbolic
Computation Group of the Theoretical Physics Department at UERJ - Brazil.}
and K. von B\"ulow\footnotemark[9] for useful discussions and a careful
reading of this paper; and Prof. M.A.H. MacCallum\footnote{Queen Mary and
Westfield College, University of London - U.K.} for kindly sending us a copy
of his paper \cite{maccallum}, as well as valuable references.

\end{document}